\documentclass{llncs}

\usepackage{graphicx} 
\usepackage{braket}
\usepackage{tikz}
\usepackage{amsmath}
\usetikzlibrary{quantikz2}

\begin{document}

\title{Fair Benchmarking of Optimisation Applications}

\author{Frank Phillipson}
%
\authorrunning{F. Phillipson}

\institute{Maastricht University, Maastricht, The Netherlands \\
TNO, The Hague, The Netherlands \\
\email{f.phillipson@maastrichtuniversity.nl}}

\maketitle

\begin{abstract}
Quantum optimisation is emerging as a promising approach alongside classical heuristics and specialised hardware, yet its performance is often difficult to assess fairly. Traditional benchmarking methods, rooted in digital complexity theory, do not directly capture the continuous dynamics, probabilistic outcomes, and workflow overheads of quantum and hybrid systems. This paper proposes principles and protocols for fair benchmarking of quantum optimisation, emphasising end-to-end workflows, transparency in tuning and reporting, problem diversity, and avoidance of speculative claims. By extending lessons from classical benchmarking and incorporating application-driven and energy-aware metrics, we outline a framework that enables practitioners to evaluate quantum methods responsibly, ensuring reproducibility, comparability, and trust in reported results.
\keywords{Quantum Computing \and Benchmarking} 
\end{abstract}

\section{Introduction}
Quantum computing is increasingly viewed as a transformative accelerator across diverse fields of science and industry. In chemistry and materials science, quantum computers could enable accurate simulation of molecular structures and reactions beyond the reach of classical methods, paving the way for new drugs and advanced materials \cite{guo2024harnessing,santagati2024drug}. In cryptography, algorithms such as Shor’s demonstrate the potential to disrupt current security protocols by factoring large integers efficiently \cite{gill2022quantum}, once suitable hardware becomes available. In artificial intelligence, quantum machine learning approaches are being explored for clustering, classification, and regression tasks \cite{gill2022quantum} and in engineering domains such as logistics, supply chain management, and transport networks, quantum methods are being investigated for optimisation of routes, schedules, and resource allocations \cite{phillipson2024quantum}.

Among these application areas, combinatorial optimisation is particularly compelling. These problems arise whenever one must select the best option from a finite but exponentially large set of possibilities, such as finding the shortest route through a set of cities (the Travelling Salesman Problem) or optimising the layout of a logistics network and the example of the engineering domains above. Because solution spaces grow factorially with problem size, even modestly sized instances quickly become intractable for brute-force methods. This computational hardness explains why industry often prefers ``good enough'' solutions quickly over exact but impractically slow ones. It was precisely this challenge that, in the 1980s and 1990s, fueled the rise of metaheuristics such as simulated annealing, tabu search, and genetic algorithms. These methods do not guarantee optimality but provide high-quality solutions within practical time frames \cite{glover2003handbook}. Today, quantum computing based optimisation is entering this same landscape, not as a replacement for exact solvers, but as a new member of the heuristic family, promising different trade-offs between quality, time, and scalability.

Alongside quantum computers, other specialised hardware platforms, such as GPUs, digital annealers, neuromorphic processors, and high-performance computing clusters, are also competing to accelerate combinatorial optimisation by exploiting their unique architectures \cite{dongarra2012high,furber2016large,matsubara2020digital}.

This diversity of paradigms presents both opportunity and challenge. Practitioners naturally want to know, for example: How do I fairly assess whether a quantum approach provides an advantage over established methods? The answer is far from straightforward. Classical benchmarking practices, developed for CPUs and GPUs, are rooted in digital complexity theory, where performance is abstracted as the number of computational steps \cite{papadimitriou1998combinatorial}. While meaningful for discrete architectures, such measures do not translate directly to analogue paradigms like quantum annealing or neuromorphic computing, where computation is governed by continuous physical processes. In these systems, runtime depends on annealing schedules, coherence times, or spiking dynamics rather than a countable sequence of operations. Without a shared and fair benchmarking approach that accounts for these differences, we risk either overstating or underestimating quantum computing’s potential.\\

\noindent Summarising, several factors make benchmarking in this field uniquely challenging:
\begin{itemize}
    \item Diverse paradigms: Quantum annealers, gate-based quantum computers, and hybrid solvers differ fundamentally in how they process information. Direct runtime comparisons can be misleading.
    \item Algorithm classes: Optimisation solvers range from exact algorithms to heuristics. Comparing a quantum heuristic with a commercial-grade problem-specific solver is often unfair unless design effort and tuning are accounted for.
    \item End-to-end workflows: Real-world performance includes not just the time spent on a (quantum) processor, but also data preparation, embedding, post-processing, and parameter tuning.
    \item Non-deterministic behaviour: Many quantum algorithms produce probabilistic outcomes, making solution quality a distribution rather than a single value.
    \item Rapidly evolving hardware: Benchmarks risk becoming obsolete quickly, and extrapolating to “future” performance is unreliable.
\end{itemize}

For these reasons, benchmarking must move beyond simplistic speed comparisons and instead embrace transparency, context, and multiple evaluation criteria. We therefore propose principles and practices for fair benchmarking of quantum optimisation algorithms, tailored for practitioners who want to evaluate and communicate performance in a responsible and transparent way.

\section{Principles for Fair Benchmarking}
Discussions about how to benchmark computing results have deep roots in the broader history of computing. As early as the 1990s, Bailey \cite{bailey} outlined “twelve ways to fool the masses” with performance results in classical supercomputing, warning against misleading comparisons. More recently, McGeoch \cite{mcgeoch2024not} has adapted those lessons to quantum computing, emphasizing the importance of reporting runtimes, tuning effort, and avoiding selective instances. Gilbert and colleagues introduced the TAQOS protocol, which extends quantum computing benchmarking beyond raw device performance to include workflow time, solution quality, and robustness across problem sets \cite{gilbert2023taqos}. Our earlier work \cite{phillipson2025fair} brought already together these development in the search for missing elements.  Other initiatives, such as Q-Score \cite{martiel2021benchmarking,van2023q}  and Qoptlib \cite{osaba2023qoptlib}, highlight the value of application-driven metrics and standardised libraries for combinatorial optimisation.

Building on this body of work, we propose the following guiding principles for benchmarking quantum optimisation algorithms in a way that is fair, transparent, and useful to practitioners:
\begin{enumerate}
    \item Measure end-to-end workflows:
Report wall-clock time from problem formulation to final solution, not just device runtime. This reflects the practical cost for users.

    \item Disclose tuning effort:
Optimisation methods often require parameter tuning. Fair benchmarking should either include tuning time in the total runtime or ensure that all solvers receive comparable tuning effort.

    \item Use comparable algorithm classes:
For example, quantum annealing should be compared with classical heuristics like simulated annealing or genetic algorithms, not with exact solvers that target different trade-offs.

    \item Use comparable hardware:
For example, the question is whether it is fair to compare a multi-million dollar quantum computer with a heuristic run on a single laptop.

    \item Report solution quality transparently:
Provide the time vs solution quality curve \cite{phillipson2022searching}, showing how solution quality evolves with time, rather than only final results. This captures important trade-offs relevant for real-world decision makers.

    \item Ensure problem diversity and reproducibility:
Use standardized, openly available benchmark sets that reflect a range of real-world structures, not just “cherry-picked” instances favourable to a given solver.

    \item Avoid speculative claims:
Do not extrapolate results to hypothetical hardware or untested scales. Report only what has been observed on existing platforms.
\end{enumerate}
Together, these principles form a framework that connects lessons learned from classical benchmarking with the unique demands of quantum and hybrid computing. They are not meant to restrict creativity in algorithm design, but to ensure that when claims are made, whether they are of speedup, efficiency, or solution quality, they are communicated in a way that others can trust, reproduce, and build upon. They set the conceptual foundation for practical benchmarking efforts.

\section{Towards Practical Benchmarking Protocols}
While Section~2 outlined the guiding principles, the community has already begun to translate them into \emph{practical benchmarking protocols}. Recent initiatives demonstrate how abstract guidelines can be operationalised into measurable criteria:

\begin{itemize}
\item \textbf{Solution quality relative to baselines:}
Metrics such as those used in Q-Score \cite{martiel2021benchmarking,van2023q} and Qoptlib \cite{osaba2023qoptlib} explicitly benchmark against best-known or optimal results. Q-Score, for example, measures the largest graph instance of Max-Cut or Max-Clique that can be solved up to a quality threshold, while Qoptlib provides standardized combinatorial optimisation problems (TSP, VRP, Bin Packing, Max-Cut) with reference solutions for fair comparison.

\item \textbf{Workflow-level timing:}  
The TAQOS protocol \cite{gilbert2023taqos} emphasizes wall-clock time for the entire workflow, including problem encoding, quantum execution, and classical post-processing. Similarly, hybrid algorithm benchmarks for QAOA and VQE explicitly include the cost of repeated quantum circuit evaluations and the time required by classical optimisers \cite{lubinski2024quantum}.  

\item \textbf{Robustness across instances:}  
Benchmarks such as Qoptlib are designed to cover families of related problems, ensuring that solvers are tested on diverse instances rather than cherry-picked cases. In addition, competitions such as the Max-Cut Challenge and cross-paradigm evaluations (e.g., comparing HPC, GPUs, and quantum devices on identical problem families \cite{van2023q}) illustrate the need for broad coverage.  

\item \textbf{Energy consumption:}  
The Quantum Energy Initiative (QEI) \cite{quantum_energy_initiative} explicitly investigates energy costs in quantum computation, addressing sustainability and life-cycle energy efficiency. This complements earlier metrics such as Quantum Volume \cite{cross2019validating}, which primarily focus on computational fidelity and scale. Incorporating energy as a benchmarking dimension is particularly relevant for analog systems (e.g., annealers, neuromorphic hardware), but also for digital quantum processors as they scale toward practical applications. TAQOS proposed to take this into the multi objective framework.
\end{itemize}

Beyond these efforts existing, we advocate a benchmarking framework that adds further dimensions:
\begin{itemize}
\item \textbf{Algorithm class awareness:} Ensure comparisons are restricted to families of algorithms with similar objectives and trade-offs.
\item \textbf{Transparent reporting:} Require disclosure of all preprocessing, parameter tuning, and hardware resources to guarantee reproducibility.
\item \textbf{Application-driven benchmarks:} Incorporate problem sets from domains such as logistics, energy, and finance to align benchmarks with real-world impact.
\end{itemize}

A particular challenge arises with \textbf{hybrid quantum--classical systems}, where we mean 'Horizontal hybrid quantum computing' following the classification by \cite{phillipson2023classification}. Algorithms such as QAOA and VQE depend on iterative feedback loops between quantum hardware and classical optimisers. Benchmarking must therefore include the full workflow: number of circuit evaluations, efficiency of classical routines, and quantum–classical communication overhead. Ignoring these aspects risks either overstating quantum contributions or underestimating integration costs. Fair protocols should capture the interaction between both components to provide a holistic picture of performance.

By moving from abstract principles to concrete protocols, benchmarking can evolve from a theoretical safeguard into a practical tool. Such a framework will help practitioners interpret performance claims in a meaningful way, while reducing the risk of overstating ``quantum advantage''.

\section{Conclusion}

Fair benchmarking is not just a technical necessity, it is vital for building trust between quantum researchers, industry stakeholders, and the public. By adopting principles of transparency, end-to-end evaluation, and fairness across algorithm classes, we can provide a more accurate picture of where quantum optimisation stands today and where it may deliver value tomorrow.

As quantum computing transitions from laboratory experiments to practical applications, robust benchmarking will be the compass that guides practitioners in deciding when, and for which problems, quantum acceleration truly matters.

\section*{Acknowledgements}
This paper was written as supporting material for the keynote presentation given at the QUEST-IS 2025 conference. This work was supported by the Dutch National Growth Fund (NGF), as part of the Quantum Delta NL program.

\bibliographystyle{splncs04}
\bibliography{biblio.bib}
\end{document}